\documentclass[12pt,epsf]{iopart}
\usepackage{graphicx}
\usepackage{amssymb,latexsym}
\usepackage{amsfonts}
\usepackage{amssymb}
\usepackage{cancel}
\usepackage{color}
\input{colordvi.tex}

\newcommand{\beq}{\begin{equation}}
\newcommand{\beqa}{\begin{eqnarray}}
\newcommand{\eeq}{\end{equation}}
\newcommand{\eeqa}{\end{eqnarray}}
\newcommand{\p}{\phi}

\newcommand{\ka}{\kappa}

\newcommand{\Om}{\Omega}
\newcommand{\hatg}{\widehat{g}}
\newcommand{\hatV}{\widehat{V}}
\newcommand{\hatp}{\widehat{\phi}}
\newcommand{\hatep}{\widehat{\epsilon}}
\newcommand{\hateta}{\widehat{\eta}}
\newcommand{\hata}{\widehat{a}}
\newcommand{\hatH}{\widehat{H}}
\newcommand{\hatt}{\widehat{t}}
\newcommand{\R}{{\cal R}}
\newcommand{\N}{{\cal N}}

\def\NPB#1#2#3{Nucl. Phys. {\bf B#1}, #2 (19#3)}

\def\PLB#1#2#3{Phys. Lett. B {\bf #1}, #2 (19#3)}

\def\PRD#1#2#3{Phys. Rev. D {\bf #1}, #2 (19#3)}

\def\PRL#1#2#3{Phys. Rev. Lett. {\bf#1}, #2 (19#3)}

\begin{document}

\title{Extended Slow-Roll Conditions and Rapid-Roll Conditions}

\author{Takeshi Chiba}%
\address{
Department of Physics, \\
College of Humanities and Sciences, \\
Nihon University, \\
Tokyo 156-8550, Japan}
\author{Masahide Yamaguchi}%
\address{Department of Physics and Mathematics, Aoyama Gakuin
University, Sagamihara 229-8558, Japan \\ and \\
Department of Physics, Stanford University, Stanford CA 94305}

\date{\today}


\begin{abstract}
We derive slow-roll conditions for a scalar field which is
non-minimally coupled with gravity in a consistent manner and express 
spectral indices of scalar/tensor perturbations in terms of the
slow-roll parameters.  The conformal invariance of the curvature
perturbation is proved without linear approximations.  Rapid-roll 
conditions are also derived, and the relation with the slow-roll conditions 
is discussed. 
\end{abstract}

\maketitle

\section{Introduction}

In curved spacetime, a scalar field may generally couple to the scalar
curvature so that the potential of $\phi$ has an additional term $\xi R
\phi^2/2$ \cite{yokoyama,fm}.  $\xi =1/6$ corresponds to the conformal
coupling: a massless scalar field is conformally invariant. Even if the
scalar field is minimally coupled with the curvature at tree level, a
non-minimal coupling is induced naturally by radiative corrections
\cite{allen}. Moreover, in supergravity theories scalar fields typically
have a curvature coupling from the K\"ahler potential. Furthermore, a
warped brane inflation in string theory involves a non-minimally coupled
scalar field \cite{kklmmt}. Thus, a non-minimal coupling of the scalar
field is well motivated from the theoretical point of view and has applications 
in cosmology (for example, Refs.~\cite{Chiba:1999wt,cy}).
Such a non-minimally coupled scalar field may also be
interesting from the observational view point. Recent observations
like the Wilkinson Microwave Anisotropy Probe (WMAP) 5 year results are
sufficiently precise to constrain inflation models severely. For
example, chaotic inflation with the quartic type of potential is
excluded at more than 95\,\% confidence level
\cite{Komatsu:2008hk}.\footnote{Such a chaotic inflation is predicted as
a simple realization of chaotic inflation in supergravity
\cite{Kadota:2007nc}.} One method to circumvent such a constraint is to
add another source of fluctuations \cite{Moroi:2005kz}. Another
interesting method is to introduce the non-minimal coupling to the
curvature, which can reduce the tensor to scalar ratio to negligible
levels \cite{Hwang1998,Komatsu:1999mt}. Then, a non-minimally coupled
chaotic inflation with a quartic type of potential is still viable. 
Thus, it is very useful to derive generalized slow-roll
conditions for such a non-minimally coupled inflaton field and provide the formulae 
of the scalar/tensor spectral indices in terms the slow-roll parameters, as in the case 
of a minimally coupled inflaton. 
However, as 
far as we know, such slow-roll conditions have not been derived in a
fully consistent manner.  A non-minimal coupling also realizes non-slow-roll 
inflation. The coupling to the curvature like $\xi R
\phi^2/2$ leads to the additional mass squared $m^2 \simeq 12 \xi H^2$
so that the inflaton cannot slow-roll with $\xi = {\cal O}(1)$. However,
in the case of the conformal coupling, it has recently been shown that the inflation
takes place with the rapidly rolling inflaton \cite{km}. 

In this paper, we first define the slow-rolling of the scalar field in
the Jordan frame and derive consistency conditions of it, which we call
extended slow-roll conditions. We then apply these conditions to several
examples. We also compute observational quantities (spectral indices of
scalar/tensor perturbations and the ratio of tensor to scalar
perturbation) and rewrite them in terms of these slow-roll
parameters. Our formulas make it possible to calculate the observational
quantities in either frame (Jordan or Einstein) using the functions appearing in the
Lagrangian only.  Next we define the rapid-rolling of the scalar field
and derive rapid-roll conditions and discuss its relation with the slow-rolling of the scalar field 
and then provide several examples.  In 
Appendix, we give several formulas in the Einstein frame which are
useful for calculations in the text and also prove the conformal invariance of the curvature 
perturbation. 

\section{Slow-Roll Inflation with a Non-minimally Coupled Scalar Field}

In this section, we derive the extended slow-roll conditions of the scalar field  which 
couples non-minimally to gravity and express scalar/tensor spectral indices in terms of these 
slow-roll parameters. 

The action in the Jordan frame metric $g_{\mu\nu}$ is 
\beq
S=\int d^4x\sqrt{-g}\left[{1\over 2\ka^2}R-F(\phi)R-
{1\over 2}(\nabla \phi)^2-V(\phi)\right].
\label{action}
\eeq
Here $\ka^2\equiv 8\pi G$ is the bare gravitational constant and
$F(\p)R$ term corresponds to the non-minimal coupling of the scalar field
to gravity.

We assume that the universe is described by the flat, homogeneous, and
isotropic universe model with the scale factor $a$. The field equations
are then given by
\beqa
&&H^2\equiv \left({\dot a\over a}\right)^2=
{\ka^2\over 3}\rho_{\p},\\
&&~~~\rho_{\p}={1\over 2}\dot\p^2+V+6H(\dot F+HF),\\
&&\dot H=-{\ka^2\over 2}\left(
\rho_{\p}+p_{\p}\right),\\
&&~~~p_{\p}={1\over 2}\dot\p^2-V-2\ddot F-4H\dot F-2F(2\dot H+3H^2),\\
&&\ddot\p+3H\dot\p+V'+6F'(\dot H +2H^2)=0,
\label{eom:scalar}
\eeqa
where the dot denotes the derivative with respect to the cosmic time 
and $V'= dV/d\p$.  The equation of motion of the scalar field
Eq.~(\ref{eom:scalar}) is also derived from the energy-momentum
conservation: $\dot\rho_{\p}+3H(\rho_{\p}+p_{\p})=0$.

\subsection{Slow-Roll Conditions}

Introducing $\Om=1-2\ka^2F$ which corresponds to a conformal factor between 
the Jordan frame and the Einstein frame, the equations of motions 
are rewritten as
\beqa
H^2\Om+H\dot\Om=\frac{\ka^2}{3}\left(\frac{1}{2}\dot\p^2+V\right),\\
\ddot\Om-H\dot\Om+2\dot H\Om=-\ka^2\dot\p^2,\label{eq:2}\\
\ddot\p+3H\dot\p+V'-\frac{3\Om'}{\ka^2}(\dot H +2H^2)=0.
\eeqa
Since under the slow-roll approximations \cite{st}, the time scale of the motion of the scalar field 
is assumed to be much larger than the cosmic time scale $H^{-1}$,  
as an extended slow-rolling of the scalar field, we assume that
$\dot\p^2\ll V, |\dot\Om|\ll H\Om$, $|\ddot\p|\ll H|\dot\p|$ 
and $|\ddot\phi| \ll |V'|$, then we obtain
\beqa
&&H^2\Om\simeq \frac{\ka^2}{3}V,\label{eq:slow:1}\\
&&3H \dot\p\simeq  -V'+6\frac{\Om'}{\ka^2}H^2\simeq 
-\Om^2\left(\frac{V}{\Om^2}\right)'
=:-V_{eff}',\label{eq:slow:2}
\eeqa
where in the second equation, we have assumed $|\dot H/H^2|\ll 1$, which should be checked later.
Note that if $V\propto \Om^2$, then $V_{eff}$ is flat and $V_{eff}'=0$ identically. 

In the following, we derive the consistency conditions for the extended
slow-rolling of the scalar field (the extended slow-roll conditions). By
computing $\ddot\p$ from Eq.~(\ref{eq:slow:2}), we obtain
\beqa
\frac{\ddot\p}{H\dot\p}\simeq -\frac{\dot H}{H^2}-\frac{V_{eff}''}{3H^2}.
\label{ratio:1}
\eeqa
Moreover, from Eqs.~(\ref{eq:slow:1}) and (\ref{eq:slow:2}) using
$\dot\Om=\Om'\dot\p$
\beqa
&&\frac{\dot\p^2}{V}\simeq \frac{\Om V_{eff}'^2}{3\ka^2 V^2},\label{ratio:2}\\
&&\frac{\dot\Om}{H\Om}\simeq -\frac{\Om'V_{eff}'}{\ka^2 V}.\label{ratio:3}
\eeqa
Hence, we finally introduce the following slow-roll parameters and 
obtain the extended slow-roll conditions:
\beqa
&&\epsilon:=\frac{\Om V_{eff}'^2}{2\ka^2 V^2};~~~~\epsilon \ll 1,\label{slow-roll:1}\\
&&\eta:=\frac{\Om V_{eff}''}{\ka^2 V};~~~~|\eta|\ll 1,\label{slow-roll:2}\\
&&\delta:=\frac{\Om'V_{eff}'}{\ka^2 V};~~~~|\delta|\ll 1,\label{slow-roll:3}
\eeqa
where we have introduced a factor of 2 in the definition of $\epsilon$
so that it accords with the standard notation of the slow-roll
parameters \cite{ll,lr}.   A useful bookkeeping rule is that a term involving 
derivatives divided $\kappa^2$ is to be treated as a small quantity.
The importance of the last condition Eq.~(\ref{slow-roll:3}) in the
background dynamics of the scalar field has not been fully
appreciated.\footnote{In \cite{Hwang:1996np,Komatsu:1999mt}, a similar
slow-roll parameter is introduced in the context of a scalar
perturbation equation.}  However, it is necessary for slow-roll
inflation both in the Jordan frame and in the Einstein frame and is
essential to relate these slow-roll parameters to the slow-roll
parameters in the Einstein frame, which are discussed in Appendix A.

We also need to make sure that 
$|\ddot\phi| \ll |V'|$ is satisfied in deriving Eq.~(\ref{eq:slow:2}).
{}From Eq. (\ref{eq:slow:2}), we have
\beqa
\frac{\ddot\p}{V'}\simeq -\frac{\dot H}{H}\frac{\dot\p}{V'}-\frac{V_{eff}''}{3H}\frac{\dot\p}{V'}
\simeq \frac{\dot H}{3H^2}\frac{V_{eff}'}{V'}+\frac{V_{eff}''}{9H^2}\frac{V_{eff}'}{V'}.
\eeqa
Therefore, comparing Eqs.~(\ref{slow-roll:2}) and (\ref{ratio:1}),
 $|\ddot\phi| \ll |V'|$ is satisfied as long as
\beqa
\left|\frac{V_{eff}'}{V'}\right|= {\cal O}(1).\label{sub:2}
\eeqa
{}Note that since from Eq.~(\ref{eq:2}) $|\dot H/H^2|$ is approximated as
\beqa
\left| \frac{\dot H}{H^2} \right| \simeq 
\left| \frac{\dot \Omega}{2H\Omega} - \frac{3\dot{\phi}^2}{2V} \right| 
\simeq 
\left| \frac{\Om'V_{eff}'}{2\ka^2 V} + \frac{\Om V_{eff}'^2}{2\ka^2 V^2}
 \right|,
\eeqa
$|\dot H/H^2|\ll 1$ is guaranteed by these conditions.

To sum up, the extended slow-roll conditions consist of three main
conditions Eqs.~(\ref{slow-roll:1}), (\ref{slow-roll:2}) and
(\ref{slow-roll:3}) and one subsidiary condition 
Eq. (\ref{sub:2}).\footnote{It is to be noted that the subsidiary condition
is a sufficient condition for slow-roll and a necessary condition is 
that  Eq.~(\ref{sub:2}) multiplied by $\epsilon, \eta$ or $\delta$ 
is sufficiently small.}

\subsection{Example: Chaotic Inflation with Non-Minimal Coupling}

As an example of extended slow-roll inflation, we consider a
non-minimally coupled scalar field for chaotic inflation with:
\beqa
V(\p)=\frac{\lambda}{n}\p^n,~~~~F(\p)=\frac{1}{2}\xi \p^2,
\label{poten:coupling}
\eeqa
where $\xi$ is a dimensionless coupling parameter and $\xi=1/6$
corresponds to the conformal coupling.

For $|\xi|\ka^2\p^2\gg 1$, from Eq.~(\ref{eq:slow:1}) $\xi<0$ is
required. Then $\Om\simeq -\xi\ka^2\p^2,\Om'\simeq -2\xi\ka^2\p,
f\simeq 1-6\xi$ and $V_{eff}'\simeq (n-4)V/((1-6\xi)\p)$ for $n \ne
4$. Hence slow-roll parameters become
\beqa
\epsilon=-\frac{(n-4)^2\xi}{2(1-6\xi)^2},~~~~\eta=-\frac{(n-4)(n-1)\xi}{1-6\xi},~~~~
\delta=-\frac{2(n-4)\xi}{1-6\xi},
\label{slow-roll:chaotic}
\eeqa
while a subsidiary condition becomes 
$|V_{eff}'/V'|=1/|1-6\xi||(n-4)/n|$. Therefore, for general $n$, $|\xi|\ll 1$
are required for slow-roll, which coincides with the conditions derived
in \cite{fm}.  An exception is the case of the quartic potential $n=4$.  In
this case from Eq.~(\ref{eq:slow:2}) for $|\xi|\ka^2\p^2\gg 1$, 
$V_{eff}'\simeq -\lambda\p/((1-6\xi)\xi\ka^2)=-4V/((1-6\xi)\xi\ka^2\p^3)$ 
and $V_{eff}'$ becomes vanishingly small, 
which corresponds to the flat plateau in the Einstein frame found by
Futamase and Maeda \cite{fm}.  The slow-roll parameters are
\beqa
\epsilon=-\frac{8}{(1-6\xi)^2\xi\ka^4\p^4},~~~~\eta=\frac{4}{(1-6\xi)\ka^2\p^2},~~~~
\delta=\frac{8}{(1-6\xi)\ka^2\p^2}.
\label{slow-roll:n=4}
\eeqa
Hence, for $n=4$ the slow-roll conditions are
automatically satisfied {\it irrespective} of $\xi$ as long as $\xi<0$,
which again coincides with the results in \cite{fm}.

On the other hand for $|\xi|\ka^2\p^2\ll 1$, $\Om\simeq 1,\Om'\simeq
-2\xi\ka^2\p, f\simeq 1$ and $V_{eff}'\simeq nV/\p$.  So slow-roll
parameters become
\beqa
\epsilon=\frac{n^2}{2\ka^2\p^2},~~~~\eta=\frac{n(n-1)}{\ka^2\p^2},
~~~~\delta=-2\xi n,
\eeqa
and $V_{eff}'/V'\simeq 1$. 
Hence $|\xi|\ll 1$ and $\ka^2\p^2\gg 1$ are required for slow-roll 
for $|\xi|\ka^2\p^2\ll 1$, which again coincides with the conditions given
in \cite{fm}.

\subsection{Perturbations}

In Appendix, it is shown that the gauge invariant curvature perturbation
$\R$ is invariant under the conformal transformation into the Einstein
frame. Then we can calculate the power spectrum $P_{\R}(k)$
\cite{ms,Hwang:1996np,Komatsu:1999mt},
\beqa
P_{\R}^{1/2}(k)=\frac{\hatH^2}{2\pi |d\hatp/d\hatt|}=\frac{H^2}{2\pi f^{1/2}|\dot\p|}
= \frac{3H^3}{2\pi f^{1/2}|V'_{eff}|}
\simeq \frac{\ka^3V^{3/2}}{2\sqrt{3}\pi \Om^{3/2}|V_{eff}'|},
\label{spec:scalar}
\eeqa
where $k$ is a comoving wavenumber at the horizon exit ($k=aH$), the
hatted variables are those in the Einstein frame and we have used
Eqs.(\ref{corres:1}) and (\ref{hatt}) and Eq.~(\ref{hath}).  In the last
equality we have assumed the slow-roll approximation.  Using 
\beqa
d\ln k=d\ln aH\simeq Hd\p/\dot\p\simeq -\frac{\ka^2 V}{\Om V_{eff}'}d\p,
\eeqa 
to the first order in the slow-roll
parameters,\footnote{Note that  $d\ln k\simeq -\ka^2\hatV d\hatp/(d\hatV/d\hatp)$ 
under the same approximations.}
the spectral index of scalar perturbation is then given by
\beqa
n_S-1\equiv\frac{d\ln P_{\R}}{d\ln k}=-6\epsilon+2\eta-3\delta.
\eeqa
Moreover, using the relation Eqs.(\ref{slowroll:e1}) and
(\ref{slowroll:e2}), we finally obtain the simple formula
\beqa
n_S-1=-6\epsilon+2\eta-3\delta=-6\hatep+2\hateta=\widehat{n_S}-1,
\label{index:scalar}
\eeqa
which proves the invariance of the spectral index under the conformal
transformation.

Tensor perturbations are also invariant under the conformal
transformation into the Einstein frame.  Then we can calculate the
tensor power spectrum $P_{h}(k)$ 
%
\beqa
P^{1/2}_h(k)=\frac{2\ka \hatH}{\sqrt{2} \pi}=\frac{2\ka H}{\sqrt{2\Om} \pi}
\simeq \frac{2\ka^2 V^{1/2}}{\sqrt{6} \pi\Om},
\label{spec:tensor}
\eeqa
where in the last equality the slow-roll approximation is assumed.  Then
the tensor spectral index is given by
\beqa
n_T\equiv \frac{d\ln P_{h}}{d\ln k}=-2\epsilon=-2\hatep=\widehat{n_T},
\label{index:tensor}
\eeqa
which is also conformally invariant.  The tensor to scalar ratio $r$ is
also calculated as
\beqa
r\equiv\frac{P_h}{P_{\R}}=16\epsilon=16\hatep=\widehat{r}.
\label{tensor-scalar}
\eeqa
Again this is also conformally invariant. Therefore, the consistency
relation for a single scalar field inflation
\beqa
r=-8n_T
\eeqa
is conformally invariant \cite{Komatsu:1999mt}. These results are summarized in Table 1. 

\begin{table}
  \begin{center}
  \setlength{\tabcolsep}{3pt}
  \begin{tabular}{|c|c| c|} 
   &Jordan frame $g_{\mu\nu}$ & Einstein frame $\hatg_{\mu\nu}=\Omega g_{\mu\nu}$
\\ \hline
 slow-roll parameters & $\epsilon,\eta,\delta$ & 
$\hatep=\epsilon,\hateta=\eta-\frac{3}{2}\delta$
 \\ 
 scalar spectral index  $n_S$& $1-6\epsilon+2\eta-3\delta$& $1-6\hatep+2\hateta$\\
tensor spectral index $n_T$& $ -2\epsilon$& $-2\hatep$\\
tensor/scalar ratio $r$ & $16\epsilon=-8n_T$& $16\hatep=-8\widehat{n_T}$\\    
  \end{tabular}
  \end{center}
\caption{Slow-roll parameters and inflationary observables in Jordan/Einstein frame}
\label{tab1}
\end{table}

It should be noted that the invariance refers to the equality between the quantities 
calculated using $V(\p)$ with $g_{\mu\nu}$ and those using $\hatV(\hatp)$ with $\hatg_{\mu\nu}$, 
{\it not} $V(\hatp)$ with $\hatg_{\mu\nu}$. 
Therefore, the observational quantities calculated by $V(\p)$ with non-minimal 
coupling are different from those by $V(\p)$ with minimal coupling. 

To demonstrate this, as an example, we calculate $n_S,n_T$ and $r$ with
Eq.~(\ref{poten:coupling}).  For $|\xi|\ka^2\p^2\gg 1$, from
Eq.~(\ref{slow-roll:chaotic}), they are calculated for $n\neq 4$
(note $\xi<0$ and $|\xi| \ll 1$)
\beqa
n_S-1=n_T\simeq(n-4)^2\xi,~~~~r\simeq-8(n-4)^2\xi.
\label{ns:nonmini:1}
\eeqa
Here we note that these are independent of the e-folding number.  
This feature can be easily understood by
calculating them in the Einstein frame. From Eqs.~(\ref{corres:1}) and
(\ref{corres:2}), the canonical scalar field $\hat{\phi}$ and the
potential $\hat{V}(\hat{\phi})$ in the Einstein frame are given by
\beqa
 \hat{\phi} &\simeq& \sqrt{\frac{1-6\xi}{|\xi|}} \frac{1}{\kappa}
                     \log{\frac{\phi}{\phi_0}}, \\ 
 \hat{V}(\hat{\phi}) &\simeq& \hat{V}_0 
            \exp{\left[\sqrt{\frac{|\xi|}{1-6\xi}}\,(n-4) \kappa \hat{\phi} \right]},
\eeqa
where $\phi_0$ is a constant field value which yields the origin of
$\hat{\phi}$ and $\hat{V}_0 = \lambda \phi_0^{n-4} / (n \xi^2
\kappa^4)$. Thus, the inflation becomes the power-law type with the
exponential potential in the Einstein frame. In fact, the slow-roll
parameters in the Einstein frame are given by
 $\hat{\epsilon} \simeq - (n-4)^2 \xi/2$ and  $\hat{\eta} \simeq - (n-4)^2 \xi$, and 
inserting them into the formulas (\ref{index:scalar}),
(\ref{index:tensor}) and (\ref{tensor-scalar}) yields the same values
as those calculated in the Jordan frame.

For $n=4$ from Eq.~(\ref{slow-roll:n=4}), $\epsilon=-1/(8\xi N^2),
\eta=1/2N$ and $\delta=1/N$, where $N$ is the e-folding number until the
end of inflation and is written for $n=4$ as
\beqa
N=\int^{t_e}_{t}Hdt\simeq \frac{(1-6\xi)}{8}\ka^2\p^2,
\eeqa
so that
\beqa
n_S-1=\frac{3}{4\xi N^2}-\frac{2}{N},~~~~n_T=\frac{1}{4\xi N^2},~~~~r=-\frac{2}{\xi N^2}.
\label{ns:nonmini:2}
\eeqa

On the other hand, for a minimally coupled ($F=0$) inflaton with the same 
potential, $n_S,n_T$ and $r$ are calculated in the standard manner \cite{lr}
\beqa
n_S-1=-\frac{n+2}{2N},~~~~n_T=-\frac{n}{2N},~~~~r=\frac{4n}{N},
\label{ns:mini}
\eeqa
which clearly shows that both (Eq.~(\ref{ns:nonmini:1}) or
Eq.~(\ref{ns:nonmini:2}) vs.  Eq.~(\ref{ns:mini})) are different. In
particular, although the tensor-scalar ratio $r$ for a minimally coupled
inflaton is generally large $r\simeq 0.13(n/2)(60/N)$ and it is so for a non-minimal coupling 
with $n\neq 4$, $r\simeq 0.32((1-n_S)/0.04)$, for a
non-minimally coupled inflaton with $n=4$ it can be small, $r\simeq
0.056(0.01/|\xi|)(60/N)^2$,  due to the extreme flatness of the effective 
potential $V_{eff}$ for $n=4$.

\section{Rapid-Roll Inflation}

Rapid-roll inflation is a novel type of inflation with a non-minimal coupling in which inflation
occurs even without slow-roll for the conformal coupling. In this
section, we derive the conditions for rapid-roll inflation for a more
general non-minimal coupling and discuss their relation with the slow-roll conditions.

Firstly, we must define rapid-roll inflation. Rapid-roll inflation is a
type of inflation without slow-roll of the scalar field. The motion of
the scalar field is determined primarily by the curvature coupling
rather than by potential term, so that the time scale of the motion is
determined by the Hubble parameter rather than by the effective mass of
the potential, so that $|\dot\p|\sim H\p$.  Nevertheless, as shown in
\cite{km}, for the conformal coupling, such a rapid motion of the scalar
field does not affect the expansion rate of the universe and the
universe inflates.  In the following, we first derive necessary conditions
of the coupling $F(\p)$ for de Sitter expansion for constant
$V(\p)=V_0$, and then derive rapid-roll conditions for general $V(\p)$.
The equation of motions are given by
\beqa
H^2=\frac{\ka^2}{3}\left(\frac{1}{2}\dot\p^2+V+6H(\dot F+HF)\right),\\
\ddot\p+3H\dot\p+V'+6F'(\dot H +2H^2)=0.
\eeqa
Then, by defining $\pi\equiv\dot\p+6HF'$, this can be written as
\beqa
H^2=\frac{\ka^2}{3}\left(\frac{1}{2}\pi^2+V+6H^2(F-3F'^2)\right),\label{eq:pi:1}\\
\dot\pi + 2H\pi +V'+(1-6F'')H\dot\p=0.\label{eq:pi:2}
\eeqa
Therefore, for constant $V=V_0$ if the following conditions are satisfied,
\beqa
F''=\frac{1}{6},~~~~~F=3F'^2, \label{cond:f}
\eeqa
then from Eq.~(\ref{eq:pi:2}) $\pi$ decays as $\pi\propto a^{-2}$ and
thus from Eq.~(\ref{eq:pi:1}) $H$ becomes constant: de Sitter expansion.
The conditions Eq.~(\ref{cond:f}) can be integrated to give
\beqa
F(\p)=\frac{1}{12}(\p -v)^2,\label{shiftedconformal}
\eeqa
where $v$ is a constant, which may be called a 'shifted conformal
coupling'.  Thus if the conditions Eq.~(\ref{cond:f}) are satisfied, even
if $\p$ itself does not move slowly, inflation occurs, which is called
rapid-roll inflation.

\subsection{Rapid Roll Conditions}

Next, we consider a general $V(\p)$ and $F(\p)$ and derive rapid-roll
conditions in terms of $V$ and $F$.  We assume that the time scale of
$\dot\pi$ is also determined by the Hubble scale, so that $\dot\pi\simeq
cH\pi$ and that the equations of motions are approximated by
\beqa
H^2\simeq \frac{\ka^2}{3}V,\label{rapid:cond:1}\\
(c+2)H\pi\simeq -V',\label{rapid:cond:2}
\eeqa
where $c$ is a proportionality constant at most of the order ${\cal
O}(1)$ and will be explicitly given later which is not given in \cite{km}. 
Note that the left hand side of Eq.~(\ref{rapid:cond:1}) does not
contain the conformal factor. Furthermore, for successful
inflation, the variation of the Hubble parameter should be slow
\beqa
  \left| \frac{\dot H}{H^2} \right| \ll 1. \label{rapid:cond:3}
\eeqa
In the following, we shall derive the
consistency conditions for rapid-roll inflation following \cite{km}. In
Eq.~(\ref{rapid:cond:1}), the first and the last terms of the right hand
side in Eq.~(\ref{eq:pi:1}) are neglected. 
Since  
\beqa
\frac{\pi^2}{2V}+6\frac{H^2}{V}(F-3F'^2)
\simeq \frac{3V'^2}{2(c+2)^2\ka^2V^2}+2\ka^2(F-3F'^2),
\eeqa
this is consistent if
\beqa
\epsilon_c:=\frac{V'^2}{2\ka^2V^2}+\frac{2(c+2)^2}{3}\ka^2(F-3F'^2);~~~~|\epsilon_c|\ll 1,
\label{rapid-roll:1}
\eeqa
Moreover, from the time derivative of Eq.~(\ref{rapid:cond:2})
\beqa
\frac{\dot\pi}{(c+2)H\pi}
\simeq -\frac{\dot H}{(c+2)H^2}-\frac{3V''}{(c+2)^2\ka^2V}-\frac{6F'V''}{(c+2)V'}.
\eeqa
Therefore, neglecting $\dot\pi-cH\pi+H(1-6F'')(\pi-6HF')$ in
Eq.~(\ref{rapid:cond:2}) is consistent if
\beqa
&&\eta_c:= \frac{V''}{\ka^2V}+\frac{2(c+2)F'V''}{V'}+\frac{c(c+2)}{3}
-\frac{c+2}{3}(1-6F'')-\frac{2(c+2)^2}{3}\frac{\ka^2(1-6F'')F'V}{V'}\\
&&|\eta_c|\ll 1,
\label{rapid-roll:2}
\eeqa
where we have assumed $|\dot H|/H^2\ll 1$, which also should be checked.
{}From the time derivative of Eq.~(\ref{rapid:cond:1})
\beqa
\left| \frac{\dot H}{H^2} \right| \simeq
\left| \frac{3}{2(c+2)} \frac{V'^2}{\kappa^2 V^2} 
       + 3 \frac{F'V'}{V} \right|.  
\eeqa
Therefore assuming $|\dot H|/H^2\ll 1$ is consistent if
\beqa
\delta_c:=\frac{V'^2}{2(c+2)\kappa^2 V^2} 
       +  \frac{F'V'}{V} ;~~~~|\delta_c|\ll 1.
\label{rapid-roll:3}
\eeqa
To sum, the rapid roll conditions consist of three conditions
Eqs.~(\ref{rapid-roll:1}), (\ref{rapid-roll:2}) and
(\ref{rapid-roll:3}).

The constant $c$ may be expressed in terms of the potential $V$ and the
coupling function $F$. 
The time derivative of Eq.(\ref{rapid:cond:1}) and Eq.(\ref{rapid:cond:2}) yields  
a quadratic equation for $c$, and  the solutions of it are given by 
\beqa
c=-\left(1 + 3\frac{F'V''}{V'}-\frac{3}{2}\frac{F'V'}{V} \right)
           \pm \sqrt{
             \left(1 - 3\frac{F'V''}{V'} +\frac{3}{2}\frac{F'V'}{V} \right)^2
       - 3\frac{V''}{\ka^2V}+\frac{3}{2}\frac{V'^2}{\ka^2V^2}
         }.
\eeqa
For $F$ satisfying Eq.~(\ref{cond:f}), using the rapid-roll conditions 
Eq.~(\ref{rapid-roll:1}) and Eq.(\ref{rapid-roll:3}), it can be approximated as 
(note that this can also be derived by setting $\eta_c\simeq 0$)
\beqa
  c \simeq - \left(1 + 3\frac{F'V''}{V'} \right)
           \pm \sqrt{
             \left(1 - 3\frac{F'V''}{V'} \right)^2
        - 3\frac{V''}{\ka^2V}
         },
\label{c:solution}
\eeqa
and for $|c|\ll 1$ it reduces to
\beqa
c\simeq  - \frac{3}{2}\frac{V''}{\ka^2V}\left(1+3\frac{F'V''}{V'}\right)^{-1}.
\eeqa
Hence $|c|\ll 1$ requires $|V''/\ka^2V|\ll 1$.  
Since $\Om=1-\ka^2\p^2/6={\cal O}(1)$, together with 
the conditions Eq.(\ref{rapid-roll:1}) and Eq.(\ref{rapid-roll:3}), this implies that 
the rapid-roll conditions are 
reduced to the slow-roll conditions, which is also seen from $|\dot\pi|\simeq cH|\pi|\ll H|\pi|$. 
Thus the relation between rapid-roll and slow-roll is clarified: $c={\cal O}(1)$ for 
rapid-roll; $|c|\ll 1$ for slow-roll. 

\subsection{Example}

As an example of rapid-roll inflation, we consider a scalar field with
the conformal coupling $F(\phi) = \phi^2/12$ and the potential
\beqa
  V(\phi) = v^4 \pm \frac12 m^2 \phi^2,
\eeqa
where $v$ is the typical energy scale of inflation and $m$ is the
inflaton mass. This type of potential often appears in hybrid inflation
(plus sign) or new inflation (minus sign).  Firstly, we derive the
rapid-roll conditions for this potential and then confirm the relation
$\dot\pi\simeq cH\pi$ by directly solving the equation of motion.

The rapid-roll parameters are estimated as
\beqa
\epsilon_c=\frac{m^4\p^2}{2\ka^2V^2},~~~~
\eta_c=\pm\frac{m^2}{\ka^2V}+\frac{(c+2)}{3}+\frac{c(c+2)}{3},
~~~~\delta_c=\frac{m^4\p^2}{2(c+2)\ka^2V^2}\pm\frac{m^2\p^2}{6V}.
\eeqa
Hence from $|\delta_c|\ll 1$ and $\epsilon_c\ll 1$, $m^2\p^2\ll
v^4$, so that $V\simeq v^4$. From $\epsilon_c\ll 1$, $m^4\p^2\ll
\ka^2v^8$.  $c$ is determined by solving $\eta_c\simeq 0$
(Eq.~(\ref{c:solution}))
\beqa
c\simeq -\frac{3}{2}\pm \frac{1}{2}\sqrt{1\mp \frac{4m^2}{H_0^2}},
\eeqa
where $H_0 \equiv \kappa v^2 / \sqrt{3}$. Then the scalar field moves
according to $\dot\pi\simeq c H\pi$ with $\pi=\dot\p+H\p$. Since $c$ is
always negative as long as the determinant is positive, $\pi \rightarrow
0$ and the inflation occurs.  Note that for new inflation (plus sign in
the determinant), we can consider the case of $m^2\gg H_0^2$, so that
one of the usual slow-roll condition, $|\eta|=|V''/\ka^2V|\ll 1$, is
violated, which is related to the situation in so-called fast-roll
inflation \cite{fast-roll}.

We can now confirm the assumption $\dot{\pi} \simeq cH\pi$ by solving
the equation of motion for the scalar field $\phi$ directly.  For
$m^2\phi^2 \ll v^4$, the equation of motion of $\phi$ is given by
\beqa
  \ddot{\phi} + 3 H_0 \dot{\phi} +(2H_0^2 \pm m^2) \phi \simeq 0.
\eeqa
Inserting $\phi = \phi_0 e^{\omega t}$ into the above equation yields
\beqa
  \omega^2 + 3H_0 \omega + 2H_0^2 \pm m^2 = 0.
\eeqa 
Then, $\omega$ is given by
\beqa
  \omega &=& -\frac32 H_0 \pm \frac12 \sqrt{H_0^2 \mp 4 m^2}=cH_0.
\eeqa
Thus, $\pi=\dot\p+H\p=(\omega+H_0)\p=(c+1)H_0\p$, and so indeed
$\dot\pi=c(c+1)H_0^2\p=cH_0\pi$.

Finally, we would like to point out that for the polynomial potential
$V(\phi) \propto \phi^n$, which often appears in chaotic inflation, one
of the rapid roll parameters $\delta_c \simeq n/6 = {\cal O}(1)$
violates the rapid-roll condition. Therefore, the rapid roll inflation
does not occur for this type of potential.

\section{Summary}

In this paper, we have derived the slow-roll conditions for the scalar
field non-minimally coupled to gravity in consistent manner, which was
made possible by rewriting the equation of motion using the conformal
factor $\Om$ and by introducing the effective potential $V_{eff}$.  The
slow-roll conditions consist of three main conditions
Eqs.~(\ref{slow-roll:1}), (\ref{slow-roll:2}) and (\ref{slow-roll:3})
and one subsidiary condition Eq. (\ref{sub:2}).  The
third condition $|\delta|\ll 1$ (Eq.~(\ref{slow-roll:3})) appears not to
have been derived before in the context of the background dynamics of
the scalar field.  However, it is necessary for slow-roll inflation both
in the Jordan frame and in the Einstein frame. These conditions enable
us to relate the slow-roll parameters in the Jordan frame
($\epsilon,\eta,\delta$) to those in the Einstein frame
($\hatep,\hateta$), so that observational quantities can be calculated
in either frame and the conformal invariance of them can be proved. We have also derived
the rapid-roll conditions by slightly generalizing those in
\cite{km}. The rapid-roll conditions consist of three conditions
Eqs.~(\ref{rapid-roll:1}), (\ref{rapid-roll:2}) and (\ref{rapid-roll:3}).
We also discussed the relation between rapid-roll and  slow-roll.

Our formulae, Eqs.~(\ref{spec:scalar}), (\ref{index:scalar}),
(\ref{spec:tensor}) and (\ref{index:tensor}), allow us to calculate the
observational quantities in either frame using the functions appearing
in the Lagrangian only. Although we have shown the conformal
invariance of the curvature perturbation beyond linear perturbation, 
the conformal invariance of the spectral indices 
of scalar/tensor perturbations is proved only in the first order in the slow-roll parameters. 
It would be interesting to extend the invariance to higher orders.

\ack

The authors would like to thank the participants of Summer Institute 2008 (Fujiyoshida, Japan, 
August 3-13, 2008), especially  Hideo Kodama, Shinji Mukohyama and Misao Sasaki for useful 
comments.
This work was supported in part by Grant-in-Aid for Scientific Research
from JSPS (No.\,17204018 and No.\,20540280 (T.C.), No.\,18740157 and No.\,19340054 (M.Y.))
and from MEXT (No.\,20040006 (T.C.)) and in part by Nihon University.

\appendix
\section{Slow-Roll Conditions and Perturbations in Einstein Frame}
\label{app1}

In this appendix, we perform the conformal transformation to the
Einstein frame and introduce slow-roll parameters and give their
relations with those in the Jordan frame. Perturbations in FRW metric are also given and 
the conformal invariance of the curvature perturbation is proved. 

Introducing
$\Om(\p)=1-2\ka^2F(\p)$, the action Eq.~(\ref{action}) is
\beqa
S=\int d^4x\sqrt{-g}\left[{\Om(\p)\over 2\ka^2}R-
{1\over 2}(\nabla \phi)^2-V(\phi)\right].
\eeqa
Introducing the Einstein metric $\hatg_{\mu\nu}$ by the conformal transformation
\beqa
\hatg_{\mu\nu}=\Om(\p)g_{\mu\nu},
\label{conf}
\eeqa
the action becomes that of a scalar field minimally coupled to Einstein
gravity
\beqa
S&=&\int d^4x\sqrt{-\hatg}\left[{1\over 2\ka^2}\widehat{R}-
\frac{1}{2\Om}\left(1+\frac{3\Om'^2}{2\ka^2\Om}
\right)(\widehat{\nabla} \phi)^2-\frac{V}{\Om^2}\right]\\
&=&\int d^4x\sqrt{-\hatg}\left[{1\over 2\ka^2}\widehat{R}-
\frac{1}{2}(\widehat{\nabla} \hatp)^2-\hatV(\hatp)\right],
\eeqa
where in the second line we have introduced a canonical scalar field
$\hatp$ with a potential $\hatV$
\beqa
d\hatp^2=\frac{1}{\Om(\p)}\left(1+\frac{3\Om'(\p)^2}{2\ka^2\Om(\p)}\right)=:\frac{f(\p)}{\Om(\p)}d\p^2\label{corres:1},\\
\hatV(\hatp)=\frac{V(\p)}{\Om(\p)^2}.\label{corres:2}
\eeqa

\subsection{Slow-Roll Conditions}

The slow-roll conditions in the Einstein frame are simply given by
\beqa
&&\hatep:=\frac{1}{2\ka^2\hatV^2}\left(\frac{d\hatV}{d\hatp}\right)^2;~~~\hatep\ll 1,\\
&&\hateta:=\frac{1}{\ka^2\hatV}\frac{d^2\hatV}{d\hatp^2};~~~~~~~~~~|\hateta|\ll 1.
\eeqa
In terms of $\p$, using Eq.~(\ref{corres:1}) and Eq.~(\ref{corres:2}),
these parameters are rewritten as
\beqa
&&\hatep=\frac{\Om V_{eff}'^2}{2\ka^2 fV^2},\\
&&\hateta=\frac{\Om^{5/2}}{\ka^2f^{1/2}V}\left(\frac{V_{eff}'}{f^{1/2}\Om^{3/2}}\right)',
\eeqa
where $V_{eff}'$ is defined in Eq.~(\ref{eq:slow:2}).  If we assign ${\cal O}(\varepsilon^2)$ 
to the slow-roll parameters ($\epsilon,\eta,\delta$),\footnote{We adopt the standard 
mathematics notation, according to which $\epsilon={\cal O}(\varepsilon^2)$ 
means that $\epsilon$ falls like $\varepsilon^2$ or faster as $\varepsilon \rightarrow 0$.}  then 
$f=1+{\cal O}(\varepsilon^2)$, and $f\simeq1$ is satisfied under the slow-roll conditions 
Eq.(\ref{slow-roll:1}) and Eq.(\ref{slow-roll:3}). 
Therefore, under the slow-roll approximations, the slow-roll parameters in the Einstein frame 
are related to those in the Jordan frame as
\beqa
&&\hatep\simeq \epsilon,\label{slowroll:e1}\\
&&\hateta\simeq \frac{\Om V_{eff}''}{\ka^2 V}- \frac{3}{2}\frac{\Om'V_{eff}'}{\ka^2 V}
=\eta-\frac{3}{2}\delta.\label{slowroll:e2}
\eeqa

\subsection{FRW metric and Perturbations}

{}From Eq.~(\ref{conf}), the line element in the Einstein frame is
\beqa
d\widehat{s}^2=\Om ds^2.
\eeqa
So, the cosmic time $\hatt$ and the scale factor $\hata$ in the Einstein frame become 
\beqa
d\hatt=\sqrt{\Om}dt,~~~~\hata(\hatt)=\sqrt{\Om(t)}a(t).
\label{hatt}
\eeqa
Hence the Hubble parameter $\hatH$ and the acceleration in the Einstein
frame become \cite{Hwang:1996np,Komatsu:1999mt,ep}
\beqa
\hatH=\frac{d\hata/d\hatt}{\hata}=\frac{1}{\sqrt{\Om}}\left(H+\frac{\dot\Om}{2\Om}\right),
\label{hath}\\
\frac{d^2\hata/d\hatt^2}{\hata}=\frac{1}{\Om}\left(\frac{\ddot a}{a}+\frac{H\dot\Om}{2\Om}+
\frac{1}{2}\left(\frac{\dot\Om}{\Om}\right)^{.}\right).
\eeqa
These show that the acceleration in the Einstein frame does not
immediately imply the acceleration in the Jordan frame. For the latter,
$|\dot\Om|\ll H\Om$ is required, which is realized by $|\delta|\ll
1$. If $|\dot\Om|\ll H\Om$, then we obtain
\beqa
&&\hatH\simeq \Om^{-1/2}H,\\
&&\frac{d^2\hata/d\hatt^2}{\hata}\simeq \Om^{-1}\frac{\ddot a}{a}.
\eeqa

Next, we consider scalar perturbations. In the longitudinal gauge
\beqa
ds^2=-(1+2\Psi(t,x^i))dt^2+a^2(t)(1+2\Phi(t,x^i))\delta_{ij}dx^idx^j,
\label{longitudinal}
\eeqa
the gauge invariant curvature perturbation, which is the curvature perturbation 
on the comoving (or velocity-orthogonal) slices,  
\beqa
\R=\Phi-\frac{H}{\dot\p}\delta\p
\label{curvature}
\eeqa
is constructed. Defining 
$\delta\Om(t,x^i)=\Om(t,x^i)-\Om(t)$ 
and noting that that $\hata(\hatt)=\sqrt{\Om(t)}a(t)$, we obtain
\beqa
\widehat{\Phi}=\Phi+\frac{\delta\Om}{2\Om(t)}.
\label{hatphi}
\eeqa
Then the invariance of $\R$ under the conformal transformation is
immediately proved using Eqs.~(\ref{hatphi}) and (\ref{hath})
\cite{ms,Hwang:1996np,Komatsu:1999mt}
\beqa
\widehat{\R}&=&\widehat{\Phi}-\frac{\hatH}{d\hatp/d\hatt}\delta\hatp\nonumber\\
&=&\Phi+\frac{\delta\Om}{2\Om}-\left(H+\frac{\dot\Om}{2\Om}\right)\frac{\delta\p}{\dot\p}\nonumber\\
&=&\Phi-\frac{H}{\dot\p}\delta\p=\R.
\eeqa
{}From Eq.~(\ref{conf}), tensor metric perturbations
$g_{ij}=a^2(\delta_{ij}+h_{ij})$ are invariant under the conformal
transformation \cite{mfb}.

Finally, we note that the conformal invariance of the curvature
perturbation is not limited to the linear perturbation and can be proved
in fully non-linear theory along the line of \cite{lms}.  In the
(3+1)-decomposition of the metric \cite{adm}
\beqa
ds^2=-\N^2dt^2+\gamma_{ij}(dx^i+\beta^i)(dx^j+\beta^j),
\eeqa
we write the three-metric $\gamma_{ij}$ as a product of the scale factor
and a perturbation $\Phi$
\beqa
\gamma_{ij}=a^2(t)e^{2\Phi(t,x^i)}.
\eeqa
In linear theory, this reduces to Eq.~(\ref{longitudinal}). Then we
define a quantity
\beqa
-\zeta\equiv \Phi-\int^{\p(t,x^i)}_{\p(t)}\frac{H}{\dot\p}d\p,
\eeqa
which reduces to $\R$ (Eq.~(\ref{curvature})) in linear theory and
coincides with $\zeta$ in \cite{lms} and is conserved on super-horizon
scales in the Einstein gravity.  Since $\hata e^{\widehat{\Phi}}=
\sqrt{\Om(t,x^i)}ae^{\Phi}$ and $\hata(\hatt)=\sqrt{\Om(t)}a(t)$, the
conformal invariance of $\zeta$ is immediately seen
\beqa
-\widehat\zeta&=&\widehat\Phi-\int^{\hatp(\hatt,x^i)}_{\hatp(\hatt)}\frac{\hatH}{d\hatp/d\hatt}d\hatp
\nonumber\\
&=&\Phi+\frac{1}{2}\ln(\Om(t,x^i)/\Om(t))-\int^{\p(t,x^i)}_{\p(t)}\frac{H}{\dot\p}
\left(1+\frac{\dot\Om}{2H\Om}\right)d\p 
\nonumber\\
&=&\Phi-\int^{\p(t,x^i)}_{\p(t)}\frac{H}{\dot\p}d\p=-\zeta.
\eeqa
$\widehat\zeta$ is a conserved quantity because it is the curvature perturbation in the Einstein frame, 
which implies the conservation of $\zeta$. 



\end{document}